# Time-dependent polarization states of high power, ultrashort laser pulses during atmospheric propagation

J. P. Palastro

*Institute for Research in Electronics and Applied Physics, University of Maryland College Park, Maryland 20740*

**Abstract**

We investigate, through simulation, the evolution of polarization states during atmospheric propagation of high power, ultrashort laser pulses. A delayed rotational response model handling arbitrary, transverse polarization couples both the amplitude and phase of the polarization states. We find that, while circularly and linearly polarized pulses maintain their polarization, elliptically polarized pulses become depolarized due to energy equilibration between left and right circularly polarized states. The depolarization can be detrimental to remote radiation generation schemes and obscures time-integrated polarization measurements.

A high power, femtosecond laser pulse propagating through atmosphere induces a time-dependent, nonlinear dielectric response through its interaction with $N_2$ and $O_2$ [1-3]. The dynamic feedback between the pulse and molecules results in several nonlinear optical phenomena, including spectral broadening, temporal compression, harmonic generation, and self-focusing [3]. Consequently, these pulses have numerous potential applications including, light detection and ranging (LIDAR), laser induced breakdown spectroscopy, directed energy beacon beams, and remote radiation generation [4-12]. In remote radiation generation, the conversion efficiency can depend sensitively on the polarization of the pulse, which can be altered by something as simple as a misaligned optic. For instance, in the two-color THz scheme, a pulse and its second harmonic with the appropriate relative phase ionize the air to create a slow directed current that drives THz radiation [7]. If the polarization vectors of two pulses are not aligned, the overall current and thus the THz yield are diminished [7]. The spectrum of supercontinuum light emitted during filamentation in argon and nitrogen has been shown to be sensitive to input polarization [8]. Finally, the nonlinear dipole moment of atmospheric constituents, which allows degenerate n-wave mixing, a source of remote harmonic generation, depends on the polarization of the pump pulse [12].

There has been a long controversy over the polarization stability of pulses during self-focusing in Kerr media [13-17]. The theoretical works typical consider idealized situations by limiting the problem to a nonlinear Schrodinger equation (NLSE), assuming pulses long enough such that the time dependence of the pulse can be neglected and the rotational response treated as instantaneous, and ignoring ionization processes [14,15]. More realistic treatments for the propagation of elliptically polarized pulses in gases have been presented by several authors examining the role of polarization in plasma filament formation [18-21]. In particular, Kolesik et al. observed that an initially elliptically polarized pulse become almost circularly polarized in the filament core in contrast with earlier work on longer pulses [14,21]. These more realistic propagation models in diatomic gases, such as atmosphere, simplify the delayed rotational response due to molecular alignment in two ways. First the rotational response is assumed to have the same electric field dependence as the electronic response [18,19]. This assumption is inconsistent with the observation that a weak probe pulse experiences -1/2 the alignment generated by a perpendicularly polarized pump [22]. Second the delayed nature of rotational response is modeled as a damped harmonic oscillator whose parameters are fit to more precise density matrix calculations of the rotational component of the molecular Hamiltonian [18.19].

There is renewed interest in the rotational dynamics of linear diatomic molecules due to its importance in atmospheric propagation as well as its potential for control of filament formation [22-27]. Characterization and understanding of atmospheric propagation of ultrashort laser pulses necessitates accurate rotational response models. Here, we implement a self-consistent, linear density matrix treatment of the rotational response for arbitrarily transverse polarized light into a propagation equation with the goal of developing a more realistic model of the polarization state dynamics during atmospheric propagation. The implemented multi-polarization rotational response eliminates the two simplifications and is consistent with the -1/2 alignment effect discussed above. The propagation equation evolves two polarization states of the electric field coupled through the delayed rotational and instantaneous electronic polarization densities. Ionization, ionization energy damping, and an isotropic plasma response are also included in the polarization density. Simulations conducted with the previously used rotational response model [18,19] and our density matrix model result in different predictions. In particular we find that elliptically polarized light appears more linearly polarized after atmospheric propagation.

We express the electric field and nonlinear polarization density vectors as envelopes modulated by a carrier wave at frequency $\omega_0$ and axial wavenumber $k$. Setting $k = k_0[1 + \delta\varepsilon(\omega_0)/2]$ where $k_0 = \omega_0/c$ and $\delta\varepsilon(\omega)$ is the shift in dielectric constant due to linear dispersion in atmosphere, and transforming to the moving frame coordinate $\xi = v_g t - z$ where $v_g = c[1 - \delta\varepsilon(\omega_0)/2]$ is the group velocity at frequency $\omega_0$, the electric field and polarization density are $\mathbf{E} = \mathbf{E}(r,z,t)e^{-ik\xi} + c.c.$ and $\mathbf{P}_{NL} = \mathbf{P}_{NL}(r,z,t)e^{-ik\xi} + c.c.$. The evolution of the transverse components of the electric field envelope are then determined by the modified paraxial equation

$$\left[\nabla_\perp^2 + 2\frac{\partial}{\partial z}\left(ik - \frac{\partial}{\partial \xi}\right) - \beta_2 \frac{\partial^2}{\partial \xi^2}\right]\mathbf{E}_\perp = 4\pi\left(ik - \frac{\partial}{\partial \xi}\right)^2 \mathbf{P}_{NL,\perp}. \quad (1)$$

where $\beta_2/\omega_0 c = (\partial^2 k/\partial \omega^2)|_{\omega=\omega_0} = 20 \ fs^2/m$ accounts for group velocity dispersion [28,29]. From here on, the subscript $\perp$ while not written is implied and refers to the left (L) and right (R) circular components of the electric field.

The nonlinear polarization density can be expressed as the sum of a free electron contribution, $\mathbf{P}_f$, and a molecular contribution, $\mathbf{P}_m$: $\mathbf{P}_{NL} = \mathbf{P}_f + \mathbf{P}_m$. The free electron polarization density includes the plasma response and a term accounting for the pulse energy lost during ionization

$$\left(ik - \frac{\partial}{\partial \xi}\right)^2 \mathbf{P}_f = \frac{1}{4\pi c^2}\left[\omega_p^2 - 8\pi c\left(ik - \frac{\partial}{\partial \xi}\right)\left(\frac{U_N v_N \eta_N + U_O v_O \eta_O}{|\mathbf{E}|^2}\right)\right]\mathbf{E}_\perp, \quad (2)$$

where $U_s$, $\eta_s$, and $\nu_s$ are the ionization potential, molecular number density, and ionization rate for specie $s$ respectively, $\omega_p^2 = 4\pi e^2 \eta_e / m_e$, $\eta_e$ is the electron number density, $e$ is the fundamental unit of charge, $m_e$ is the electron mass, and the subscripts N and O refer to molecular nitrogen and oxygen. The densities evolve according to $\partial_\xi \eta_e = \nu_N \eta_N + \nu_O \eta_O$ and $\partial_\xi \eta_s = -\nu_s \eta_s$.

In general the ionization rate is a function of the ellipticity, $\varepsilon^2 = 4|E_L||E_R|/(|E_L|+|E_R|)^2$, of the electric field. Perelomov et al. (PPT) derive separate ionization rates for linear and circular polarized fields [30], but a computationally efficient ionization rate, spanning both the multi-photon and tunneling regimes, and handling arbitrary ellipticity is lacking. To account for arbitrary ellipticity, we calculate the ionization rate by performing a quadratic interpolation in the ellipticity between the linear polarized and circularly polarized ionization rates: $\nu_s = \varepsilon^2 \nu_{s,l} + (1-\varepsilon^2)\nu_{s,c}$ where $\nu_{s,l}$ and $\nu_{s,c}$ are the linear and circularly polarized ionization rates provided in Ref. [30] as Eqs. (43), (54), and (68) with the Coulomb correction presented in Ref. [31]. The quadratic interpolation ensures a continuous, differentiable ionization rate at the transition from right to left circular polarization. In calculating $\nu_s$, we use $U_N = 15.6\ eV$, $Z_N = 0.9$, $U_O = 12.1\ eV$ and $Z_O = 0.53$ to match the experimental results of Talebpour *et al.* for molecular ionization [32].

The molecular contribution to the polarization density is the product of an effective nonlinear molecular susceptibility and the vector electric field: $\mathbf{P}_m = (\ddot{\chi}_{el} + \ddot{\chi}_{rot})\mathbf{E}$, where $\ddot{\chi}_{el}$ is the instantaneous electron susceptibility tensor and $\ddot{\chi}_{rot}$ the delayed rotational susceptibility tensor. The diagonal and off-diagonal electronic susceptibility tensor elements are given respectively by $(\ddot{\chi}_{el})_{aa} = \varpi_{el}(|E_a|^2 + 2|E_b|^2)$, and $(\ddot{\chi}_{el})_{LR} = (\ddot{\chi}_{el})_{RL}^* = 0$, where $a$ and $b$ refer to R and L, $\varpi_{el} = (1/6\pi^2 \eta_{atm})\sum_s \eta_s n_{2,s}$, $\eta_{atm} = 2.6 \times 10^{19}\ cm^{-3}$, $\eta_N = 0.8\eta_{atm}$ and $\eta_O = 0.2\eta_{atm}$ upstream from the laser pulse, and $n_{2,N} = 7.4 \times 10^{-20}\ cm^2/W$ and $n_{2,O} = 9.5 \times 10^{-20}\ cm^2/W$, experimentally measured values [12,33].

The rotational susceptibility tensor is found from the linear in intensity, density matrix solution for a thermal gas of linear diatomic molecules experiencing a torque in the presence of a laser electric field. The torque, proportional to the anisotropy in the molecular polarizability parallel and perpendicular to the principle molecular axis, aligns the molecules along the laser pulse polarization axis (see supplemental material). Based on the rotational degrees of freedom of the full molecular Hamiltonian, the molecules are modeled as rigid rotors with quantized angular momenta and field free energy eigenvalues $E_j = \hbar^2 j(j+1)/2I_M$, where $j$ is the total angular momentum quantum number and $I_M$ is the moment of inertia, $8.8\times10^{-28}$ and $1.2\times10^{-27}$ $eV \cdot s^2$ for nitrogen and oxygen respectively. For an arbitrarily transverse-polarized electric field, the rotational susceptibility tensor elements for a gas specie, $s$, can be written as a sum over susceptibility contributions from each total angular momentum state: $(\ddot{\chi}_{rot,s})_{ab} = \sum_j (\ddot{\chi}_{j,s})_{ab}$. In the following, we leave off the specie subscript for simplicity, but note that the simulations solve for the susceptibility contributions of $O_2$ and $N_2$ independently. The susceptibility contributions, $(\ddot{\chi}_j)_{ab}$, evolve according to the following equation:

$$\left[\frac{d^2}{d\xi^2} + \frac{\omega_{j,j-2}^2}{c^2}\right](\ddot{\chi}_j)_{ab} = -Q_j F_{ab}, \quad (3)$$

where $F_{aa} = F_{bb} = (|E_a|^2 + |E_b|^2)/3$, $F_{ab} = 2E_a E_b^*$ for $a \neq b$,

$$Q_j = \frac{\eta(\Delta\alpha)^2}{5\hbar} \frac{j(j-1)}{2j-1}\left(\frac{\rho_{j,j}^0}{2j+1} - \frac{\rho_{j-2,j-2}^0}{2j-3}\right)\omega_{j,j-2},$$

$\Delta\alpha = \alpha_\parallel - \alpha_\perp$, $\alpha_\parallel$ and $\alpha_\perp$ are the linear polarizabilities along and perpendicular to the molecular bond axis respectively, $\omega_{j,j-2} = \hbar(2j-1)/I_M$, $\rho_{jj}^0 \equiv \sum_m \langle m|\rho_{ab}^0|m\rangle$, $\rho_{ab}^0 = \delta_{ab} Z_p^{-1} D_j \exp[-E_j/T]$, $T$ is the temperature, $D_j$ a degeneracy factor associated with nuclear spin, and $Z_p$ the partition function $Z_p = \sum_j (2j+1)D_j \exp[-E_j/T]$. For $\Delta\alpha$ the experimentally measured values, $\Delta\alpha_N = 7\times10^{-25} cm^3$ and $\Delta\alpha_O = 1.1\times10^{-24} cm^3$, are used [33]. For comparison, extending the previous used rotational response [18,19] with density matrix theory results in $F_{aa} = (8/3)(|E_b|^2 + 2|E_a|^2)$, $F_{bb} = (8/3)(|E_a|^2 + 2|E_b|^2)$, and $F_{ab} = 0$ for $a \neq b$. These couplings will be used in comparisons of the polarization evolution.

The off-diagonal elements in the rotational susceptibility tensor are complex, providing a mechanism for energy exchange between the L and R polarization states. The exchange requires a relative phase between the polarization states that varies along the pulse. This can occur in the absence of time dynamics [$\partial_\xi \to 0$ in Eq. (1)] because the states undergo different amounts of self and cross-phase modulation. Inclusion of the time dynamics in Eq. (1) ensures the correct group velocity for the spectral components resulting from phase modulation. The changes in group velocity reshape the intensity profile of each state which alters the phase modulation, further modifying the relative phase and consequently the energy exchange.

To illustrate the interaction between the polarization states, we take $\nabla_\perp \to 0$ in Eq. (1), assume $k \gg \partial_\xi$, set $E_a = |E_a|\exp(i\phi_a)$, and obtain the following equations for the energy density of each mode and the relative phase:

$$\frac{\partial}{\partial z}|E_L|^2 = 4\pi k |E_R||E_L| \operatorname{Im}\left[(\tilde{\chi})_{LR} e^{-i\Delta\phi}\right], \quad (4a)$$

$$\frac{\partial}{\partial z}\left(\frac{\Delta\phi}{2\pi k}\right) = -\varpi_{el}[|E_L|^2 - |E_R|^2] - \operatorname{Re}\left[\frac{|E_L|^2 (\tilde{\chi})_{RL} e^{i\Delta\phi} - |E_R|^2 (\tilde{\chi})_{LR} e^{-i\Delta\phi}}{|E_L||E_R|}\right], \quad (4b)$$

$$(\tilde{\chi}_j)_{RL} = -\sum_{s,j}\left(\frac{2Q_j}{\omega_{j,j-2}}\right)\int_{-\infty}^{\xi} \sin\left[\frac{\omega_{j,j-2}(\xi-\xi')}{c}\right]|E_R||E_L|e^{-i\Delta\phi}d\xi', \quad (4c)$$

and $\partial_z|E_R|^2 = -\partial_z|E_L|^2$, where $\Delta\phi = \phi_L - \phi_R$. From Eq. (4a), we see that if $(\tilde{\chi}_j)_{RL} = 0$, there is no energy exchange between the two polarization states. Furthermore, if $\Delta\phi$ is independent of $\xi$, $\operatorname{Im}[(\tilde{\chi})_{LR} e^{-i\Delta\phi}] = 0$ and $\partial_z|E_L|^2 = \partial_z|E_R|^2 = 0$: the amplitude of each polarization state does not evolve. Equation (4) also demonstrates that pure circularly or pure linearly polarized pulses maintain their polarization. For circular polarization, either $|E_R| = 0$ or $|E_L| = 0$, and the RHS of Eq. (4a) is zero, while for linear polarization, $|E_L|^2 = |E_R|^2$, $\Delta\phi = \pi/2$, and $\operatorname{Im}[(\tilde{\chi})_{LR} e^{-i\Delta\phi}] = 0$. Note that in the absence of an off diagonal susceptibility, Eq. (4) disallows energy exchange between the polarization states: the susceptibility included here provides a fundamentally different interaction between the polarization states than that used previously [18,19]. By neglecting the time dynamics ($k \gg \partial_\xi$), Eq. (4) does not capture the energy gain and loss associated with spectral shifting in the time dependent susceptibility. The rate at which a state's energy changes due to spectral shifting depends on the amplitude of both states, a consequence of cross phase modulation. Thus, when time dynamics are included, spectral shifting provides an energy transfer mechanism between the states that does not require an off diagonal susceptibility. This energy exchange, however, requires a temporal variation in the enveloped amplitude or susceptibility nearing the laser period to be comparable to the mechanism described by Eq. (4).

We simulate the laser pulse evolution by solving Eq. (1) in azimuthally symmetric, cylindrical coordinates. The propagation of the pulse is simulated over a distance of 5.5 $m$ starting from a focusing lens with a 3 $m$ focal length and $f_\# = 590$. The initial transverse profile of each pulse is Gaussian with an initial waist of 0.26 $cm$ and a vacuum focal waist of 300 $\mu m$. The initial longitudinal intensity profile is $\sin^4(\pi \xi / \sigma)$ for $0 < \xi < \sigma$, with $\sigma = 139$ $fs$. The corresponding FWHM is $\sigma_{FWHM} = 50$ $fs$. To characterize the polarization of the pulse, we use the normalized, spatially averaged Stokes parameters defined as follows: $S_0 = \int [|E_L|^2 + |E_R|^2] d^2r d\xi$, $S_1 = 2 S_0^{-1} \int |E_L||E_R| \cos(\Delta\phi) d^2r d\xi$, $S_2 = 2 S_0^{-1} \int |E_L||E_R| \sin(\Delta\phi) d^2r d\xi$, and $S_3 = S_0^{-1} \int [|E_L|^2 - |E_R|^2] d^2r d\xi$, where $\Delta\phi = \phi_L - \phi_R$. The degree of polarization can be written in terms of the Stokes parameters as $d = (S_1^2 + S_2^2 + S_3^2)^{1/2}$. The third stokes parameter provides a convenient indication of the polarization: for a left or right circularly polarized pulse $S_3 = \pm 1$ respectively and for a linearly polarized pulse $S_3 = 0$. The degree of polarization diagnoses the variability of the polarization: if the polarization is distinctly linear, elliptical, or circular at every point within the pulse, $d = 1$, and if the pulse is completely depolarized $d = 0$. The radial integration for calculating the Stokes is performed over the entire simulation domain, 0.72 $cm$.

In Figure (1) $S_3$ and $d$ are plotted as a function of initial value, $S_{3,i}$, at three distances from vacuum focus $-50$ (red), $-18$ (green), and 100 $cm$ (blue) for an initial pulse energy of 1 mJ. Figure (1a) displays $S_3$ and $d$ for the delayed rotational response implemented here. For $S_{3,i} = 0$ and $S_{3,i} = 1$, the value of $S_3$ changes little during propagation, while for $0 < |S_{3,i}| < 1$, $S_3$ decreases significantly: a small ellipticity results in the time averaged polarization evolving away from circular polarization towards linear polarization. The decrease in degree of polarization is largest for pulses that have undergone the largest changes in $S_3$, suggesting variability of the polarization state within the pulse.

Figure (1b) highlights the difference between our rotational response model based on the molecular Hamiltonian and the model assuming the instantaneous and rotational susceptibilities have identical electric field dependence. Figure (1b) displays $S_3$ and $d$ for the latter model. As opposed to our model, $S_3$ remains constant during propagation. This constancy is expected: as discussed above, the lack of off-diagonal elements in the susceptibility nearly eliminates energy transfer between circular states. The pulse does, however, become depolarized. In the linearly polarized basis, the susceptibility of the previous model has off-diagonal elements, allowing transfer of energy between linearly polarized states.

Returning now to the model presented here, the variation of the polarization states within the pulse, characteristic of depolarization, is demonstrated in Fig. (2). Figure (2) displays the power in the L (blue) and R (red) polarization states as a function of the moving frame coordinate initially, left, and 1 $m$ after vacuum focus, right for $S_{3,i} = 0.976$ (a) and $S_{3,i} = 0.22$ (b). When $S_{3,i} = 0.976$, the less energetic R-state becomes amplified by the L-state at the back of the pulse. The rate of energy transfer increases from the front of the pulse backwards consistent with the delayed temporal response associated with molecular alignment. Because of this shorter pulses, $\sigma_{FWHM} \sim 25$ $fs$, may maintain their polarization over longer distances.

The polarization state varies from L-circular at the front of the pulse, to elliptical, linear, elliptical, and finally linear again at the back of the pulse. The inversion in power near the back of the pulse suggests that the states undergo an energy oscillation in an attempt to equilibrate. Through refractive and diffractive spreading, however, the electric field amplitudes drop, the exchange weakens, and one state is left more energetic. For atmospheric propagation with larger $f_\#$ and over longer distances, we expect the polarization states to undergo additional oscillations. For $S_{3,i} = 0.22$, the polarization states nearly equilibrate except at the front of the pulse where the molecular alignment is minimal.

Figure (3) shows $S_3$ as a function of initial value, $S_{3,i}$, at three distances from vacuum focus −50 (red), −18 (green), and 100 $cm$ (blue) for an initial pulse energy of 3 mJ. The inset shows the degree of polarization at the same distances. Similar to the 1 mJ case, $S_3$ drops from −50 to −18 $cm$. However, an increase in $S_3$ occurs between −18 and 100 $cm$ most noticeably for $S_{3,i} = 0.82$: the less energetic R-state is transferring energy to the more energetic L-state. This can occur when the electric field amplitude of the R-state is larger than that of the L-state in regions where the coupling between the states is strongest, regions of high intensity/fluence. This is illustrated in Fig. (4) showing the total fluence, color scale, and the on-axis fluence of the L (blue) and R (red) polarization states as a function of propagation distance for $S_{3,i} = 0.65$, $S_{3,i} = 0.82$, and $S_{3,i} = 0.94$. For $S_{3,i} = 0.65$ and $S_{3,i} = 0.94$ the R-state fluence is bounded above by that of L-state, while for $S_{3,i} = 0.82$ the R-state fluence surpasses the L-state at $z = 0.15\ m$. This inversion causes a transfer of energy from the less energetic R-state to more energetic L-state.

We have investigated the evolution of the polarization states for high power femtosecond laser pulses propagating through atmosphere. To calculate the effective rotational susceptibility, density matrix theory was applied to the rotational degrees of freedom in the molecular Hamiltonian. The Hamiltonian included the external potential of the arbitrary transverse-polarized laser electric field. The resulting susceptibility tensor possessed off-diagonal terms allowing energy exchange between circular polarization states. The susceptibility model corrects a common misassumption: that the rotational susceptibility has the same symmetry as the instantaneous susceptibility [18,19]. This misassumption underestimates the energy transfer between circular states and as demonstrated here strongly affects the polarization evolution. The simulations predict that initially circular or linearly polarized pulses maintain their polarization, while initially elliptically polarized pulses become depolarized during atmospheric propagation. The depolarization was the result of energy transfer between polarization states mediated by the rotational response. The polarization may be increasingly modified during atmospheric propagation over longer distances due an extended interaction between the states. Furthermore, because of the delayed nature of the rotational response, shorter pulses may be more polarization stable.

The author would offers thanks to J. Wahlstrand, H.M. Milchberg, T.M. Antonsen, P. Sprangle, W. Zhu, L. Johnson, T. Rensink and C. Miao. This work was supported by ONR, NSF, and DOE.

## References

[1]  A. Braun et al., Opt. Lett. **20**, 73 (1995).
[2]  P. Sprangle et al., Phys. Rev. E **66**, 046418 (2002).
[3]  A. Couairon and A. Mysyrowicz, Phys. Rep. 441, 47 (2007).
[4]  G. Mejean et al., Appl. Phys. B **78**, 535 (2004).
[5]  B. Kearton and Yvette Mattley, Nat. Photonics **2**, 537 (2008).
[6]  P. Sprangle et al., Proc. SPIE, 8535 (2012).
[7]  K.Y. Kim et al., Opt. Express **15**, 4577 (2007).
[8]  O. Varela et al., Opt. Express **17**, 3630 (2009).
[9]  Y. Liu et al. Opt. Commun. **284**, 4706 (2011).
[10]  A.J. Traverso et al. Proc. Natl. Acad. Sci. **109,** 15185 (2012).
[11]  J. Penano et al., J. Appl. Phys. **111**, 033105 (2012).
[12]  G. Agrawal, *Nonlinear Fiber Optics* (Academic Press 2007)
[13]  D. H. Close et al., IEEE J. Quant. Electron. **2**, 553 (1966).
[14]  J. H. Marburger, Prog. Quant. Electron. **4**, 35 (1975).
[15]  G. Fibich and B. Ilan, Phys. Rev. E **67**, 036622 (2003).
[16]  A.H. Sheinfux et al., Appl. Phys. Lett. **101**, 201105 (2012).
[17]  L. Arissian et al., arXiv :1206.6933v1 (2012).
[18]  M. Kolesik et al. Phys. Rev. E. **64**, 046607 (2001).
[19]  A. Couairon et al., Opt. Comm. **225**, 177 (2003).
[20]  O. Kosareva et al. Opt. Lett. **35**, 2904 (2010).

[21] N.A. Panov et al. Quantum Electron. **41**, 160 (2011).
[22] Y.-H. Chen et al., Opt. Express **15**, 11341 (2007).
[23] J.P. Palastro et al., Phys. Rev. A **84**, 013829 (2011).
[24] J.P. Palastro et al., Phys. Rev. A **85**, 043843 (2012).
[25] J.P. Palastro et al., Phys. Rev. A **86**, 033834 (2012).
[26] S. Varma et al. Phys. Rev. Lett. **101**, 205001 (2008).
[27] S. Varma et al., Phys. Rev. A **86**, 023850 (2012).
[28] E.R. Peck and K. Reeder, J. Opt. Soc. Am. **62**, 958 (1972).
[29] P. Wrzesinksi et al. Opt. Express **19**, 5163 (2011).
[30] A.M. Perelomov et al., JETP **23**, 924 (1966).
[31] S.V. Popruzhenko et al., Phys. Rev. Lett. **101**, 193003 (2008).
[32] A. Talebpour et al., Opt. Commun. **163**, 29 (1999).
[33] J. K. Wahlstrand et al. Phys. Rev. A **85**, 043820 (2012).

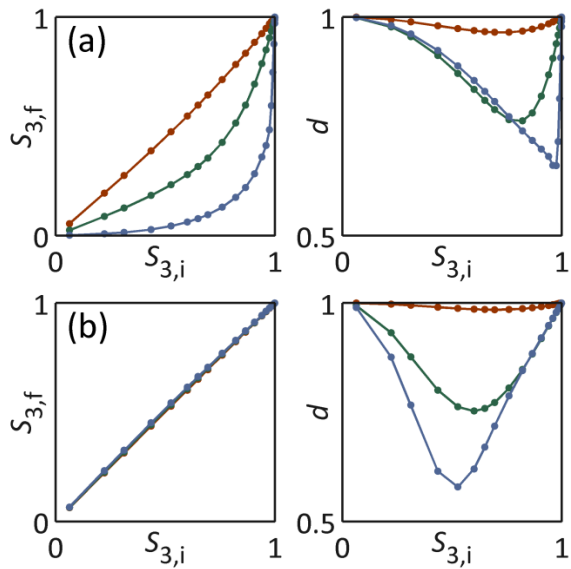

Fig. 1. value of $S_3$, left, and $d$, right, as a function of initial $S_3$ value for a pulse energy of 1 mJ at distances from vacuum focus of $-50$ (red), $-18$ (green), and 100 $cm$ (blue). (a) results when using the using the delayed susceptibility derived from density matrix theory applied to the rotational degrees of freedom in the molecular Hamiltonian. A slight ellipticity causes $S_3$ to drop substantially resulting in depolarization. (b) results when using a delayed susceptibility with the same amplitude dependence as the instantaneous

susceptibility. $S_3$ remains unchanged. The depolarization results from a drop in $S_1^2 + S_2^2$ (not shown).

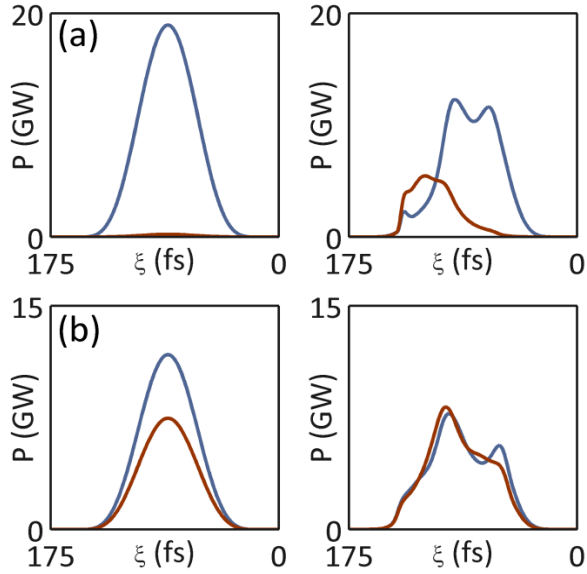

Fig. 2. Power in the L circular (blue) and R-circular (red) polarization states as a function of moving frame coordinate at $-150\ cm$, left, and $100\ cm$, right, before and after vacuum focus, respectively. (a) $S_3 = .976$ initially. (b) $S_3 = .22$ initially. The more energetic L-state transfers power to the R-state during propagation. The delayed nature of the rotational response results in larger energy transfer at the back of the pulse.

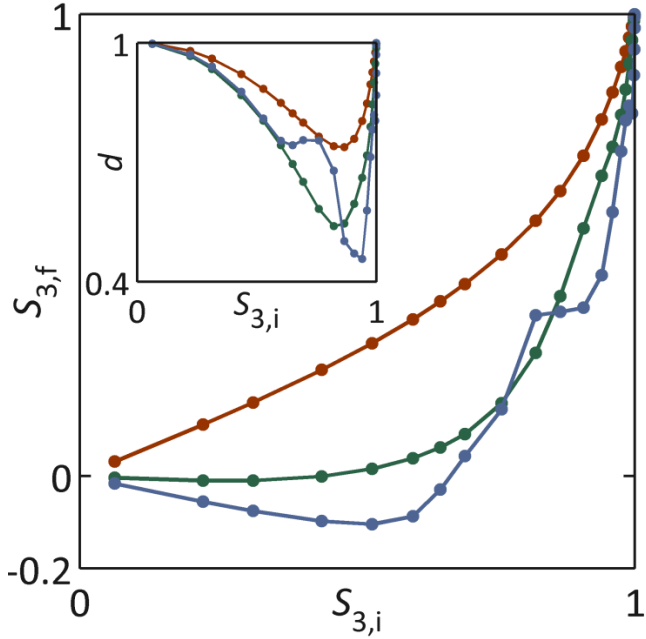

Fig. 3. value of $S_3$ as a function of initial $S_3$ value for a pulse energy of 3 mJ at distances from vacuum focus of $-50$ (red), $-18$ (green), and $100$ cm (blue). The degree of polarization at the same distances is displayed in the inset. $S_3$ drops initially but increases between $-18$ and $100$ cm for a range of initial $S_3$ values, most noticeably $S_{3,i} = 0.82$. For intermediate values of $S_{3,i}$ the degree of polarization increases after initially dropping: the polarization is initially elliptical and uniform within the pulse, the ellipticity then varies within the pulse, and finally the polarization relaxes to a linearly polarized state and increased values of $d$.

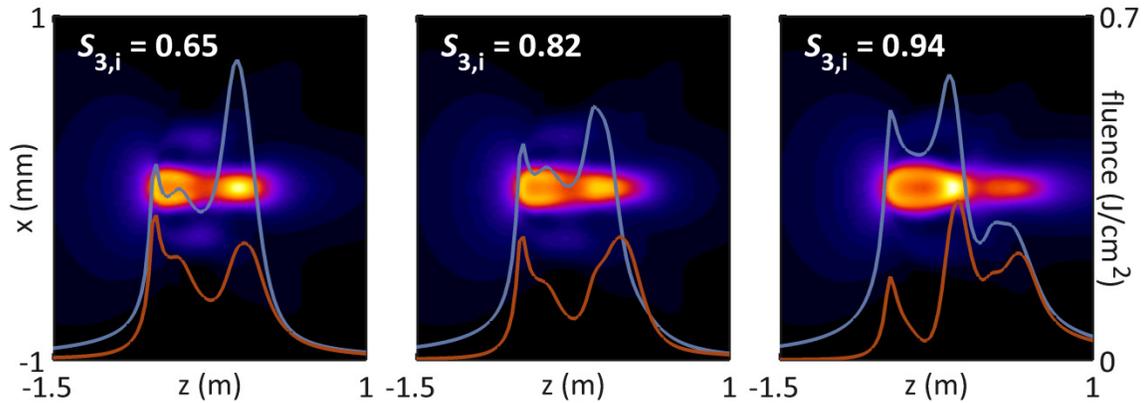

Fig. 4. total fluence as a function of transverse coordinate and distance from vacuum focus for three initial $S_3$ values. The lines show the on-axis fluence for the L-circular (blue) and R-circular (red) polarization states. For $S_{3,i} = 0.65$ and $S_{3,i} = 0.94$ the on axis fluence of the R-state is bounded above by the L-state. For $S_{3,i} = 0.82$ the on-axis fluence of the R-state surpasses that of the L-state at $z = 0.15\ m$. The inversion in fluence results in a local transfer of energy from the less energetic R-state to the more energetic L-state, resulting in the growth of $S_3$ observed in Fig. 3.